\documentclass[12pt,flushrt]{article}
\usepackage{aaspp4}
\begin{document}

\title{Alignments of the Dominant Galaxies in Poor Clusters}
\author{Todd M. Fuller\altaffilmark{1} and Michael J. West\altaffilmark{2}}

\affil{Department of Astronomy and Physics, Saint Mary's University, Halifax
NS, B3H 3C3, Canada}

\authoremail{tfuller@phobos.astro.uwo.ca, west@ap.stmarys.ca}

\and
\author{Terry J. Bridges\altaffilmark{2}}
\affil{Institute of Astronomy, Cambridge, UK}
\authoremail{tjb@ast.cam.ac.uk}

\altaffiltext{1}{present address: Department of Physics and Astronomy, 
University of Western Ontario,
London ON, N6A~3K7, Canada}

\altaffiltext{2}{Visiting Observer, Jacobus Kapteyn Telescope.  The JKT
is operated on
the island of La Palma by the Isaac Newton Group in the Spanish
Observatorio del Roque de los Muchachos of the Instituto de
Astrofisica de Canarias}

\begin{abstract}

We have examined the orientations of brightest cluster galaxies (BCGs)  in poor
MKW and AWM clusters and find that, like their counterparts  in richer Abell
clusters, poor cluster BCGs exhibit a strong   propensity to be aligned with
the principal axes of their host clusters  as well as the surrounding
distribution of nearby $(\leq 20\,h^{-1}$ Mpc) Abell clusters. The processes
responsible for dominant galaxy alignments are therefore independent of cluster
richness.   We argue that these alignments  most likely arise from anisotropic
infall  of material into clusters along large-scale filaments.

\end{abstract}


\section{Introduction} 

The orientation of galaxies is one more piece to be fit into the puzzle of
galaxy formation.   Statistically significant evidence for alignments between
the principal axes of rich Abell clusters and the major axes of their dominant
galaxies (hereafter referred to as Brightest Cluster Galaxies, or BCGs) has
been reported by numerous authors (Sastry 1968; Carter \& Metcalfe 1980;
Struble \& Peebles 1985; Rhee \& Katgert 1987; Lambas, Groth, \& Peebles
1988).  Struble (1990) and Trevese, Cirimele, \& Flin (1992)  found that the
BCG major axis is also aligned with the line joining the first and second
brightest galaxies, and that the second brightest galaxy is weakly aligned with
the first.  There is also solid evidence that  BCGs and their parent clusters
are aligned with the distribution of neighbouring clusters on  scales up to
several tens of Mpc\renewcommand{\thefootnote}{\fnsymbol{footnote}}
\footnote[2]{We adopt H$_0$=100 $h\,$ km/sec/Mpc in this paper}
\renewcommand{\thefootnote}{\arabic{footnote}} (Binggeli 1982; Lambas et al.
1990; West 1994).  For instance, West (1994) finds that there is significant
alignment (at $>$99.9\% confidence) of the innermost regions ($\leq$ 2h$^{-1}$
kpc) of 147 Abell Cluster BCGs with the distribution of neighbouring rich
clusters out to 10h$^{-1}$ Mpc.

While many studies have been done, they have focused almost exclusively on rich
clusters,  so the effect of cluster environment on the alignment effects is not
known.  In order to investigate whether cluster richness influences the
alignment effects, we have obtained CCD images of the BCGs of some  poor
clusters.  Morgan, Kayser, \& White (1975, MKW) and later Albert, White, \&
Morgan (1977, AWM)  catalogued 23 candidate BCGs located in poor clusters. 
These poor clusters contain a few tens of bright galaxies and have virial
masses of $10^{13}$--$10^{14}$ M$_\odot$ (e.g., Beers et al. 1995), compared to
$10^{14}$--$10^{15}$ M$_\odot$  for rich Abell clusters (e.g., Carlberg et al.
1996).  Beers et al. (1995) find a median velocity dispersion of 336 km/sec for
21 MKW/AWM poor clusters, about half that   found in rich clusters  (e.g.,
Zabludoff et al. 1990).

Flin et al. (1995) examined the MKW and AWM poor clusters for alignments
between the parent cluster position angle and the position angles of the two
brightest galaxies and found an alignment for the first brightest galaxies but
not the second.  Their observations were taken in 1986 with a 105 cm Schmidt
telescope using photographic plates, and galaxy and cluster orientations were
estimated by eye.  Our study improves considerably on this previous work, since
CCDs are much more sensitive and have improved linearity over photographic
plates and thus capture faint features more reliably.  Also, our use of
automated surface photometry procedures allow a more accurate determination of
galaxy position angles. Using these data, we have investigated whether poor
cluster BCGs are aligned with their host cluster, and if they are also aligned
on larger scales with the distribution of surrounding Abell clusters.  

Our observations and data reductions are described in the following section. In
\S III we examine the evidence for alignments of BCGs in  poor clusters, and in
\S IV we discuss the theoretical implications  of our results.

\section{Observations and Data Reduction}

Images of BCGs in 21 of the 23 MKW/AWM clusters were obtained using the 1.0m
Jacobus Kapteyn Telescope (JKT) during an observing run in April 1994 and
another in April 1995. With the JKT in its $f/15.0$ configuration,  the image
scale of $0''{\!\!.}33$ and CCD size of $1124 \times 1124$ yielded a field of
$6'{\!\!.}2 \times 6'{\!\!.}2$.  We  obtained 900 sec $V$ band images in the
first run and 900 sec $B$ and/or 600 sec $R$ images in the second.

\begin{figure}[h]

\plottwo{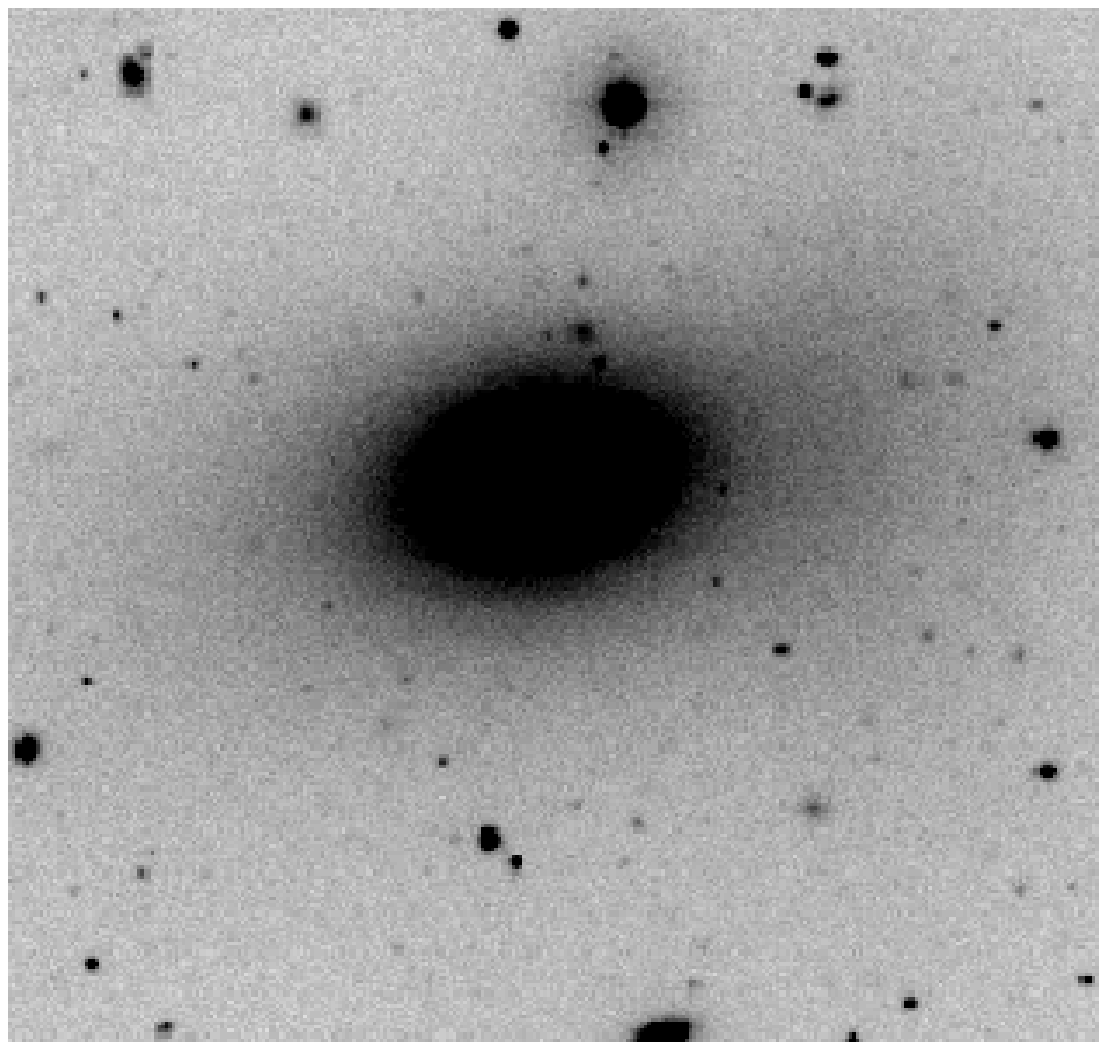}{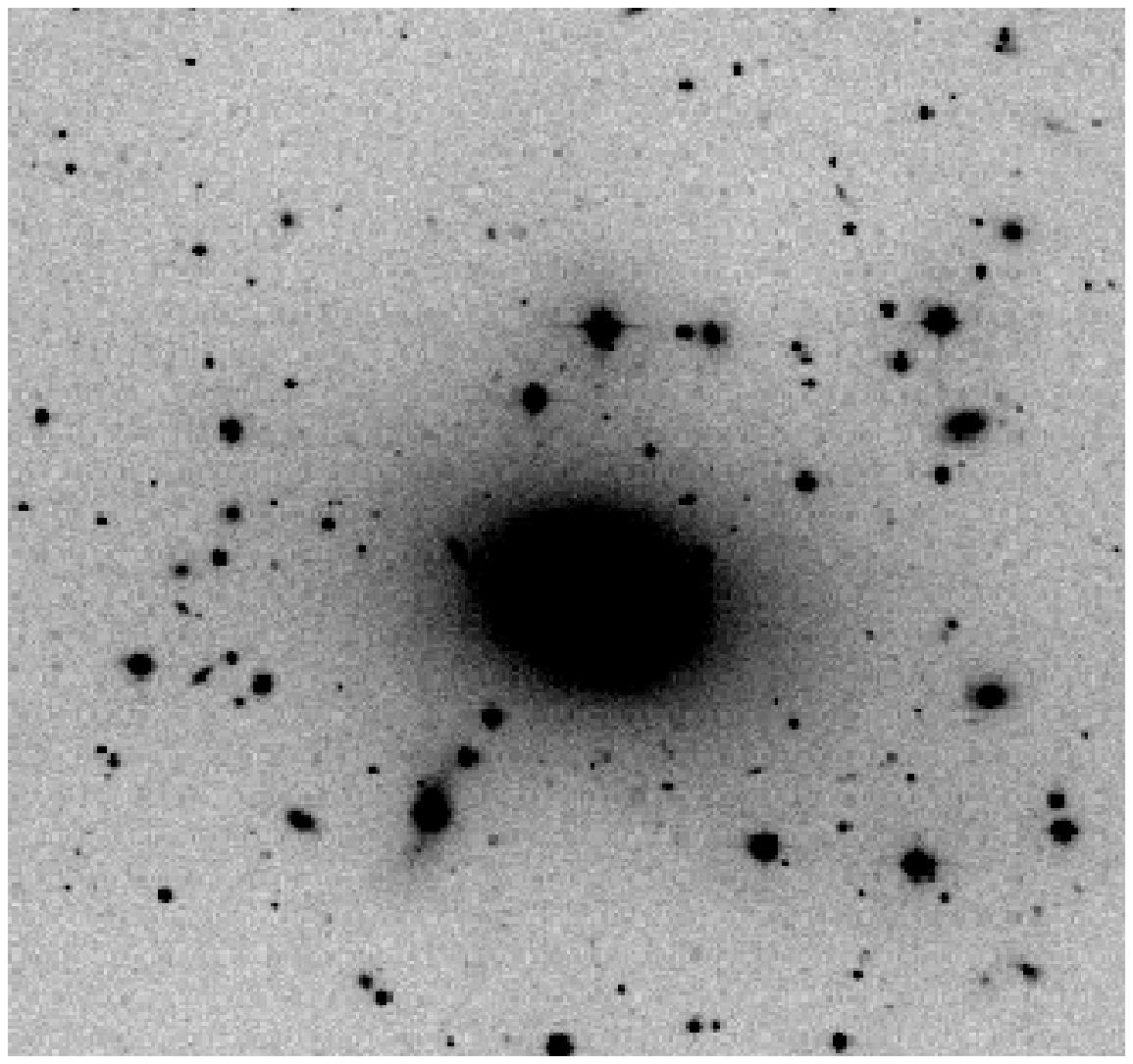}

\caption{JKT B images of two poor cluster BCGs:~  
{\it left:} MKW 4; {\it right:} AWM 5
\label{figimages}}

\end{figure}

The data were preprocessed (bias-subtracted, flat fielded, and trimmed) using
the {\ttfamily IRAF CCDPROC} package.  Flat fielding was performed using both
twilight flats and dark sky flats, and the residual gradients in the final
images were about 1.0\% of the sky intensity.  Contaminating objects (e.g.
stars, cluster and background galaxies) were identified by eye and masked using
generous radii.  Figure \ref{figimages} shows images of two galaxies in our
sample.

The BCG position angles were measured with the {\ttfamily STSDAS} task
{\ttfamily ellipse}.  This task uses the iterative method  of Jedrzejewski
(1987) to fit isophotal ellipses to galaxies.  The user supplies initial
estimates for the position angle, ellipticity, and ellipse center, and
specifies the final semi-major axis distance.  The routine samples the images
along an elliptical path and produces a one dimensional intensity distribution
as a function of the ellipse eccentric anomaly, $E$.  The Fourier harmonics of
the distribution are fit by least-squares to the function  $$y=y_0+A_1\sin(E) +
B_1\cos(E)+A_2=\sin(2E)+B_2\cos(2E).$$  Next, the five ellipse parameters are
adjusted by a correction found from the amplitudes $A_1, B_1, A_2$, and $B_2$. 
The parameter with the largest amplitude is varied, a new elliptical path is
chosen, and the image is resampled.  The task stops after a user specified
number of iterations or after the solution has converged, and the best fitting
ellipse is given by the parameters that produced the lowest absolute values of
the harmonic amplitude.  The output consists of the five ellipse parameters
($x$ and $y$ centroids, ellipticity, position angle and axis length), plus
higher-order harmonics characterizing the departures from purely elliptical
isophotes.  The routine is fairly insensitive to the initial estimates; the
deviations incurred here are much smaller than the variation in the position
angle with radius.  

\section{Analysis}

\begin{figure}[h]

\plotfiddle{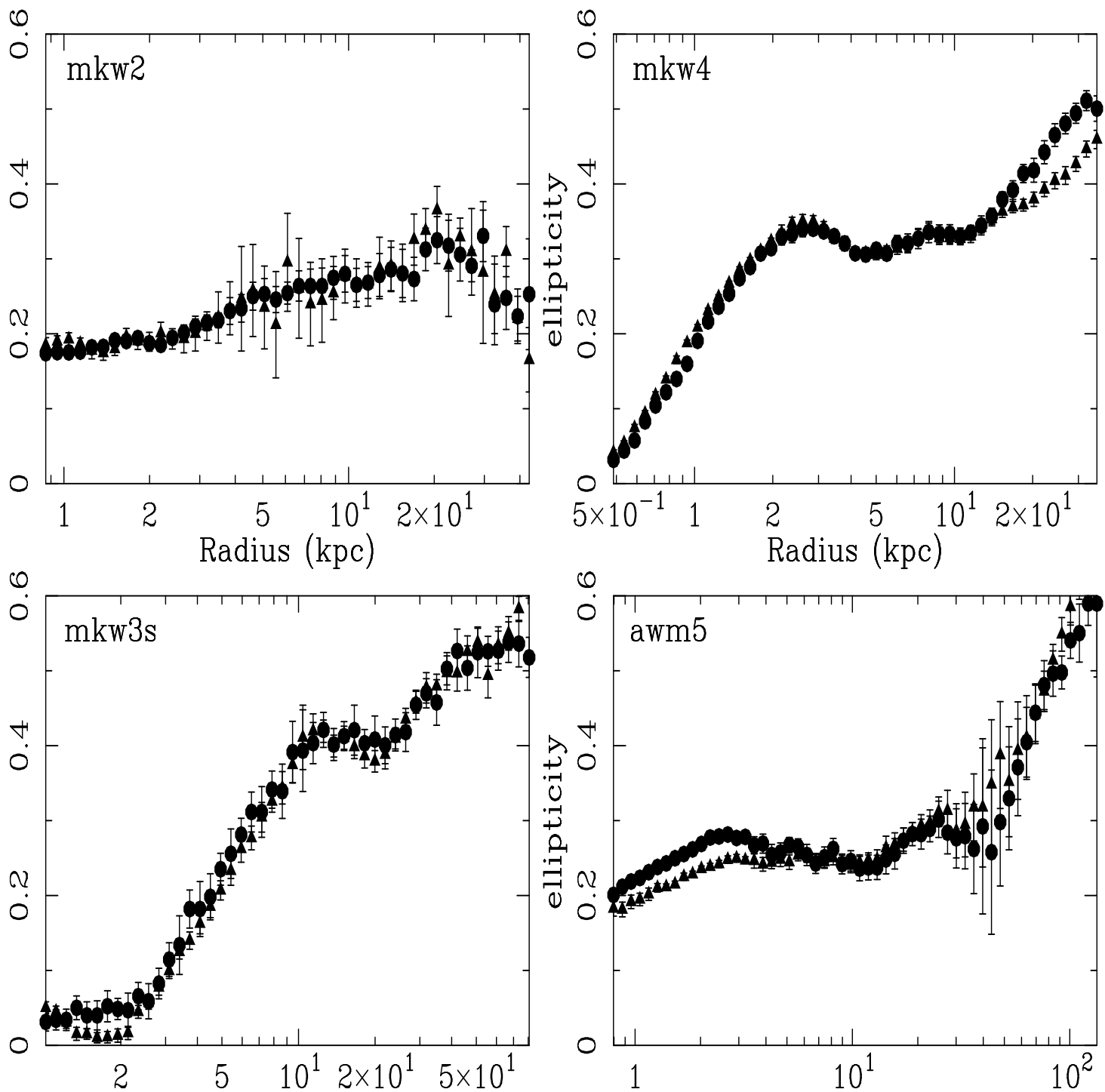}{7.0 truein}{270}{75}{100}{-300}{575}

\caption{Ellipticity versus radius (kpc) profile measured in $B$ (circles) and
$R$ (triangles) for a sample of the AWM/MKW poor cluster BCGs.  
\label{figepsprofile}}

\end{figure}

\begin{figure}[h]

\plotfiddle{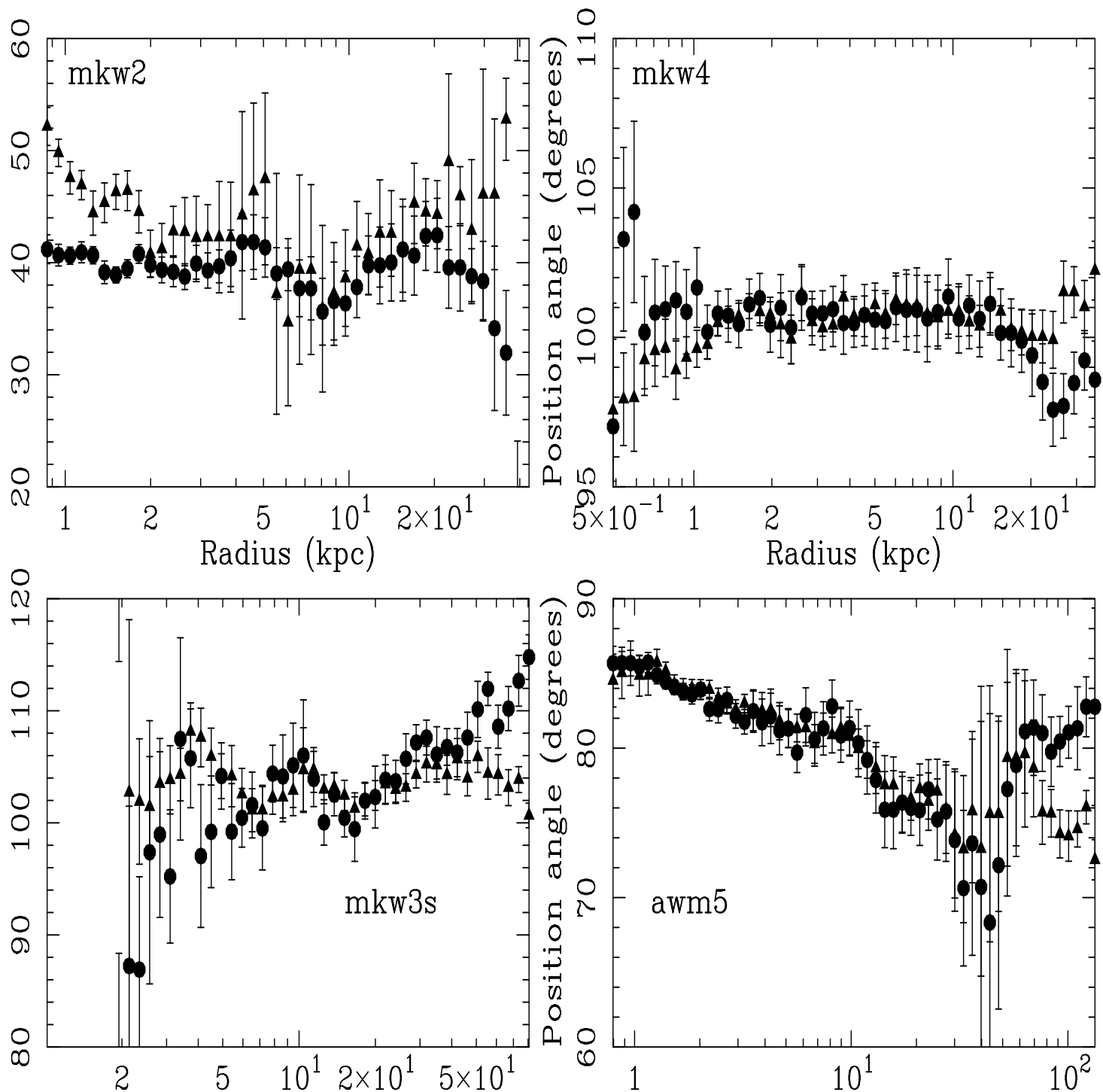}{7.0 truein}{270}{75}{100}{-300}{575}

\caption{Position angle versus radius (kpc) profile measured in $B$ (circles) and
$R$ (triangles) for a sample of the AWM/MKW poor cluster BCGs.  
\label{figpaprofile}}

\end{figure}

The position angles listed in Table \ref{tabalignangle} were averaged over
radii less than 6 kpc where the isophote intensity was high.  The errors given
are one standard deviation of the mean.  The central
few arcseconds were excluded since seeing effects become stronger in the core
and tend to make the isophotes round.  For 5 galaxies (MKW 2, 2S, 3S, 4, and
AWM 5) we obtained both $B$ and $R$ data, and the position angles given are
average values.  Figures \ref{figepsprofile} and \ref{figpaprofile} show the
ellipticity and position angle profiles for a sample of the poor cluster BCGs.
None of the galaxies in this study displayed significant isophotal twisting. 
The largest degree of twisting was $\sim 20^\circ$ in AWM 3, which is much
smaller than the $40^\circ$ twists found by Porter et al. (1991) in a sample of
bright ellipticals.

\begin{figure}[h]

\plotfiddle{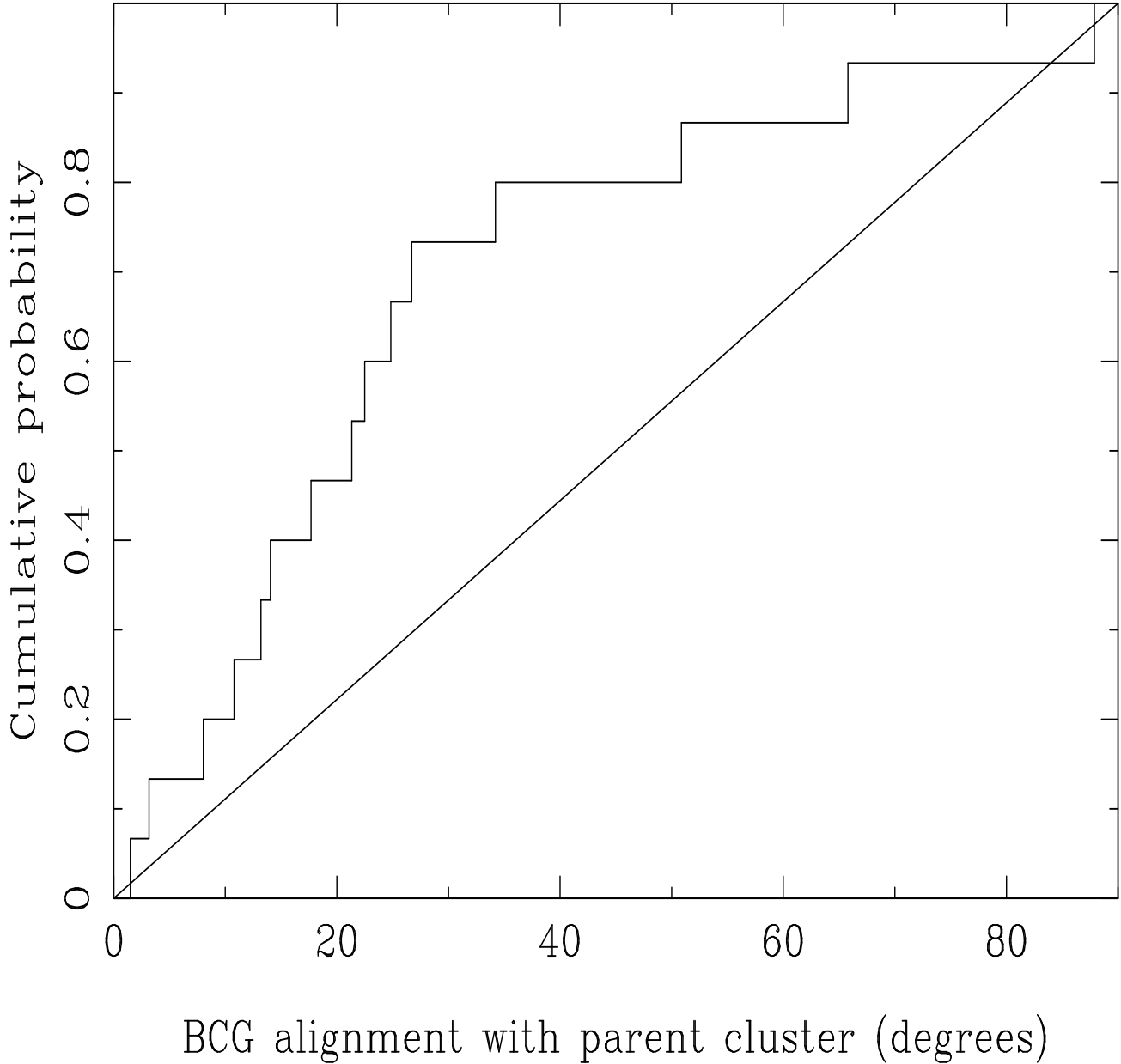}{7.0 truein}{270}{75}{100}{-300}{575}

\caption{Cumulative probability distribution of the angle between BCG
major axis and host cluster principal axis; the straight line is that
expected for a random distribution.  The K-S test shows that
poor cluster BCGs are aligned within their host clusters at the 
92\% confidence level.
\label{fighostkstest}}

\end{figure}

To determine if the BCG position angles were aligned with their parent
clusters, we applied the Kolmogorov-Smirnov (K-S) test.    We discarded any
galaxies with mean position angle errors of $ \geq 15^\circ$, or mean
ellipticities ($1-b/a$) of less than 0.2 since round galaxies do not have well
defined major axes.   After application of these limitations there were 9
BCG-parent cluster pairs. Figure \ref{fighostkstest} shows the  cumulative
probability distribution as a function of the alignment angle between the BCGs
and their parent clusters, and for a random distribution of angles.  The
cluster position angles were taken from Flin et al. (1995).  Since the
alignment angles must be sorted according to size to apply the K-S test and
random angles are all equally probable, the cumulative probability distribution
for a random set of angles is a straight line with a slope of 1.  The BCG
position angles clearly show a departure from a random distribution.  The K-S
test allows us to reject the hypothesis that the alignment angles come from a
random distribution at the 92\% level.  Thus, we may say with confidence that
the major axes of poor cluster dominant galaxies are aligned with their parent
clusters, in agreement with Flin et al. (1995).

\begin{figure}[h] 

\plotfiddle{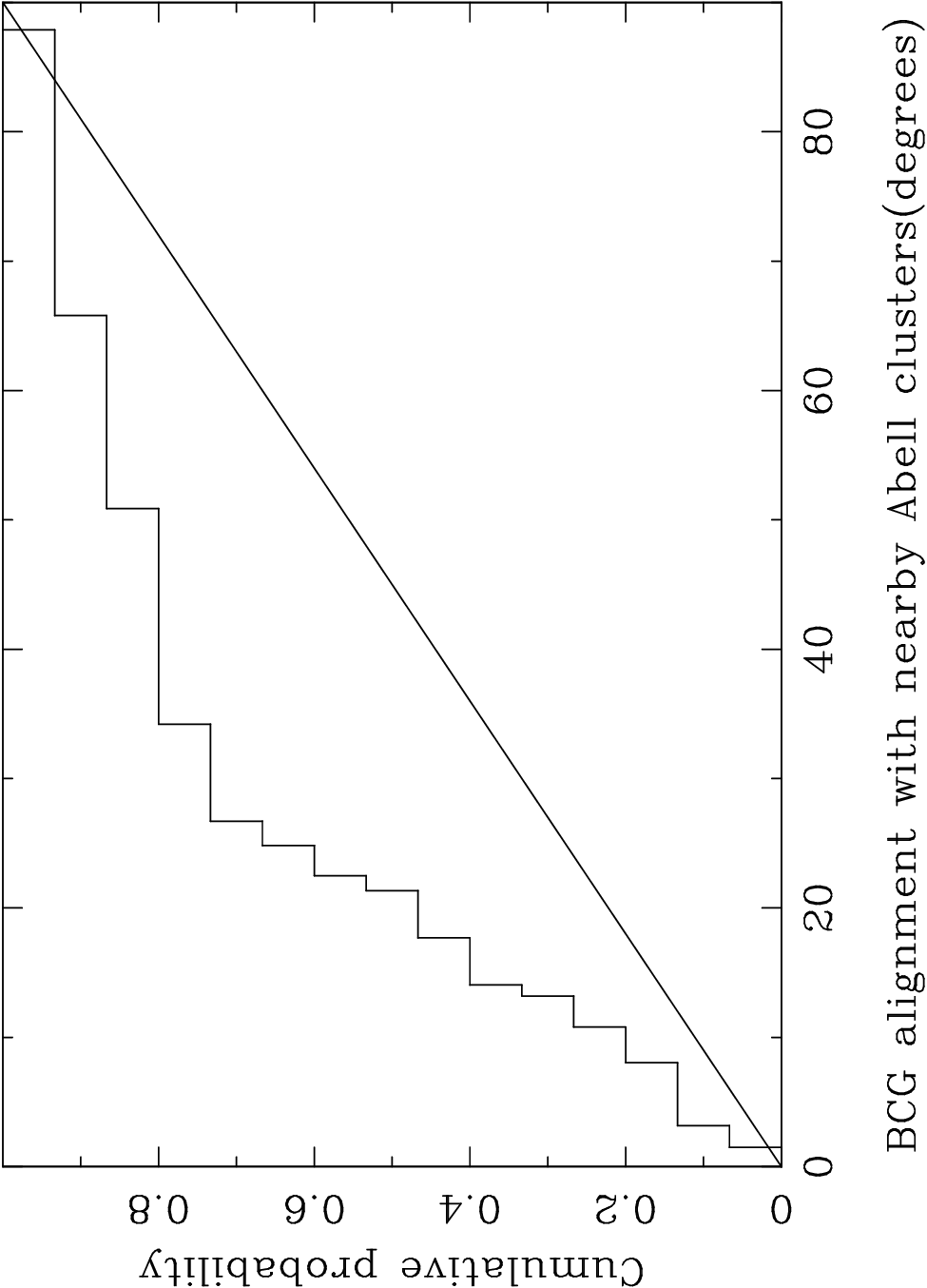}{7.0 truein}{270}{75}{100}{-300}{575}

\caption{Cumulative probability distribution of the acute angle between the BCG
major axis and the great circle connecting the galaxy with a neighboring
Abell cluster within $20\,h^{-1}$ Mpc.
The observed distribution 
(stepped line) shows a clear departure from a random distribution
(straight line) at the 99\% confidence level. \label{figkstest}}

\end{figure}

Next, we located the Abell clusters within $20h^{-1}$ Mpc and measured the
acute angle between the BCG major axis and the great circle connecting the
galaxy with the Abell cluster.  After applying the above restrictions on the
ellipticity and position angle, our sample contained 10 MKW/AWM clusters that
had one or more neighbouring Abell clusters; in all there were 15 galaxy-Abell
cluster pairs.  Figure \ref{figkstest} shows the cumulative probability
distribution for the angle between the BCG major axis and the principal axis of
the host cluster.  The K-S test shows that the distribution is non-random at
the 99\% level.   Hence these results indicate that   BCGs in poor clusters are
as strongly aligned with their environs as are  their counterparts in richer
Abell clusters (see, e.g., West 1994).

The K-S test does not take into consideration the errors in the position
angles, so we performed a series of 1000 K-S tests on data generated by adding
random deviations of $\pm 15^\circ$ to the measured position angles.  This
value was chosen since it is one of our criteria for rejection of position
angles and is a generous error allowance.  The median significance level of the
1000 K-S tests for the BCG-parent cluster data was $92.3\% \pm 8\%$, and
$98.9\% \pm 1\%$ for the BCG-Abell cluster data (errors are $\pm 1 \sigma)$.  
This robust test demonstrates that the alignment effect is not easily masked by
variations of up to $15^\circ$ in the position angles.

A K-S test using only those Abell clusters within $10h^{-1}$ Mpc was
inconclusive owing to the small sample size, as there are  relatively few
neighboring clusters within this separation.  When all Abell clusters within
$30h^{-1}$ Mpc were included no significant departures from  randomness were
found.

\begin{deluxetable}{rrrrr}
\tablecolumns{3}
\tablewidth{0pc}
\tablecaption{Position angles \label{tabalignangle}}
\tablehead{
\colhead{cluster}		& 
\colhead{$\theta_{BCG}$\tablenotemark{a}} 	&
\colhead{BCG ellipticity} &
\colhead{$\theta_{cluster}$\tablenotemark{b}} \\
& \colhead{(degrees)}
&
& \colhead{(degrees)}}
\startdata
MKW 1  &  46 $\pm$ 3  & 0.28  &  50\\
MKW 2  &  41 $\pm$ 2  & 0.22  &  60\\
MKW 4  & 101 $\pm$ 1  & 0.30  & 120\\
MKW 5  &  98 $\pm$ 5  & 0.13  \\
MKW 6  & 106 $\pm$ 2  & 0.45  &  56\\
MKW 7  & 168 $\pm$ 3  & 0.10  & 150\\
MKW 8  & 103 $\pm$ 11 & 0.14  &  10\\
MKW 9  &  55 $\pm$ 30 & 0.05  \\
MKW 10 & 171 $\pm$ 4  & 0.40  & 175\\
MKW 11 & 172 $\pm$ 8  & 0.25  & 165\\
MKW 12 & 105 $\pm$ 7  & 0.14  &  65\\
MKW 1s &  15 $\pm$ 1  & 0.24  & 140\\
MKW 2s & 148 $\pm$ 2  & 0.10  & 130\\
MKW 3s & 103 $\pm$ 1  & 0.21  &   \\
MKW 4s &  30 $\pm$ 1  & 0.38  &  55\\
AWM 1  &  65 $\pm$ 2  & 0.21  &  20\\
AWM 3  &  92 $\pm$ 1  & 0.26  \\
AWM 4  & 167 $\pm$ 1  & 0.25	\\
AWM 5  &  82 $\pm$ 1  & 0.26	\\
AWM 6  & 116 $\pm$ 4  & 0.45  \\
AWM 7  &  69 $\pm$ 5  & 0.20	\\

\enddata
\tablenotetext{a}{Position angles were measured north through east.}
\tablenotetext{b}{Cluster position angles from Flin et al. 1995}
\end{deluxetable}

\section{Discussion}

The observed alignments of BCGs, clusters and superclusters $-$ a  coherence of
structures over scales from tens of kpc to tens of Mpc $-$ must  surely be an
important clue about how these objects formed. The results presented in this
paper  provide an important new piece of information: whatever mechanism is
responsible for  producing alignments of BCGs with their surroundings, it
appears  to operate equally well in both rich and poor clusters.  

We believe that these alignments are readily explained by  hierarchical models
of structure formation in which  BCGs and clusters are built by infall of  
material that flows along the filamentary superclusters in which they  are
embedded.   In such a scenario   clusters and their brightest member galaxies
are   built by a series of mergers that occur preferentially  along the
direction  defined by the filament, and hence these objects  will naturally
develop  orientations that reflect the surrounding filamentary  pattern of
superclustering.  In this way the matter  distribution on supercluster scales
influences the properties  of clusters and their BCGs. Such a process would be
expected to produce alignments of BCGs in both poor and rich clusters.

This picture of cluster and BCG formation via anisotropic mergers  is strongly
supported  by theoretical work and numerical simulations which have shown that 
infall of material into clusters along filaments is a  generic feature of most
gravitational instability models of structure formation. (e.g., Bond 1987;
Bond, Kofman \& Pogosyan 1996;   van Haarlem \& van de Weygaert 1993; West
1994; Dubinski 1998). Observational evidence also supports this idea; for
example,  West, Jones \& Forman (1995) showed that the distribution  of merging
subclusters in clusters -- the building blocks from which  rich clusters are
made -- traces the surrounding  filamentary distribution of matter on
supercluster scales. Assuming that BCGs formed by mergers, then it is natural
to  expect that such mergers will also occur preferentially along the 
direction defined by the cluster principal axis, which is itself  dictated by
the surrounding filamentary mass distribution on  supercluster scales.

The fact that BCGs in poor clusters exhibit the same alignment  effect that is
seen in the richer Abell clusters indicates that this  alignment phenomenon is
not limited to the most massive galaxy clusters. The possibility that such
galaxy alignments might extend to even sparser  groups is worth exploring.

\section{Summary}

We have shown that the brightest member galaxies in poor MKW/AWM clusters  are
preferentially aligned with the principal axes of their host clusters, and that
they also point towards nearby rich clusters.    BCG-parent cluster alignments
and BCG-nearby cluster alignments are observed in both poor and rich clusters,
and furthermore the degree of alignment is very significant in both types of
clusters.  These two observations assert that cluster richness cannot be a
factor  in producing alignments.  We suggest that these alignments are  most
likely produced by formation of (rich and poor) cluster BCGs by infall along
filamentary structures, which are a generic feature of many models for the
formation of large scale structure.

\section{Acknowledgments}

MJW and TMF were supported by a grant from NSERC of Canada.

\clearpage 

\clearpage

\clearpage

\clearpage

\clearpage

\end{document}